\documentclass[twocolumn,showpacs,preprintnumbers,amsmath,amssymb,prc,superscriptaddress]{revtex4}
\usepackage{graphicx}
\usepackage{dcolumn}
\usepackage{bm}

\begin{document}

\title{Multiplicity and Pseudorapidity Distributions of  Charged Particles and Photons at Forward Pseudorapidity in Au + Au Collisions at $\sqrt{s_{\mathrm {NN}}}$ = 62.4 GeV}

\medskip

\affiliation{Argonne National Laboratory, Argonne, Illinois 60439}
\affiliation{University of Bern, 3012 Bern, Switzerland}
\affiliation{University of Birmingham, Birmingham, United Kingdom}
\affiliation{Brookhaven National Laboratory, Upton, New York 11973}
\affiliation{California Institute of Technology, Pasadena, California 91125}
\affiliation{University of California, Berkeley, California 94720}
\affiliation{University of California, Davis, California 95616}
\affiliation{University of California, Los Angeles, California 90095}
\affiliation{Carnegie Mellon University, Pittsburgh, Pennsylvania 15213}
\affiliation{Creighton University, Omaha, Nebraska 68178}
\affiliation{Nuclear Physics Institute AS CR, 250 68 \v{R}e\v{z}/Prague, Czech Republic}
\affiliation{Laboratory for High Energy (JINR), Dubna, Russia}
\affiliation{Particle Physics Laboratory (JINR), Dubna, Russia}
\affiliation{University of Frankfurt, Frankfurt, Germany}
\affiliation{Institute of Physics, Bhubaneswar 751005, India}
\affiliation{Indian Institute of Technology, Mumbai, India}
\affiliation{Indiana University, Bloomington, Indiana 47408}
\affiliation{Institut de Recherches Subatomiques, Strasbourg, France}
\affiliation{University of Jammu, Jammu 180001, India}
\affiliation{Kent State University, Kent, Ohio 44242}
\affiliation{Lawrence Berkeley National Laboratory, Berkeley, California 94720}
\affiliation{Massachusetts Institute of Technology, Cambridge, MA 02139-4307}
\affiliation{Max-Planck-Institut f\"ur Physik, Munich, Germany}
\affiliation{Michigan State University, East Lansing, Michigan 48824}
\affiliation{Moscow Engineering Physics Institute, Moscow Russia}
\affiliation{City College of New York, New York City, New York 10031}
\affiliation{NIKHEF and Utrecht University, Amsterdam, The Netherlands}
\affiliation{Ohio State University, Columbus, Ohio 43210}
\affiliation{Panjab University, Chandigarh 160014, India}
\affiliation{Pennsylvania State University, University Park, Pennsylvania 16802}
\affiliation{Institute of High Energy Physics, Protvino, Russia}
\affiliation{Purdue University, West Lafayette, Indiana 47907}
\affiliation{Pusan National University, Pusan, Republic of Korea}
\affiliation{University of Rajasthan, Jaipur 302004, India}
\affiliation{Rice University, Houston, Texas 77251}
\affiliation{Universidade de Sao Paulo, Sao Paulo, Brazil}
\affiliation{University of Science \& Technology of China, Hefei 230026, China}
\affiliation{Shanghai Institute of Applied Physics, Shanghai 201800, China}
\affiliation{SUBATECH, Nantes, France}
\affiliation{Texas A\&M University, College Station, Texas 77843}
\affiliation{University of Texas, Austin, Texas 78712}
\affiliation{Tsinghua University, Beijing 100084, China}
\affiliation{Valparaiso University, Valparaiso, Indiana 46383}
\affiliation{Variable Energy Cyclotron Centre, Kolkata 700064, India}
\affiliation{Warsaw University of Technology, Warsaw, Poland}
\affiliation{University of Washington, Seattle, Washington 98195}
\affiliation{Wayne State University, Detroit, Michigan 48201}
\affiliation{Institute of Particle Physics, CCNU (HZNU), Wuhan 430079, China}
\affiliation{Yale University, New Haven, Connecticut 06520}
\affiliation{University of Zagreb, Zagreb, HR-10002, Croatia}

\author{J.~Adams}\affiliation{University of Birmingham, Birmingham, United Kingdom}
\author{M.M.~Aggarwal}\affiliation{Panjab University, Chandigarh 160014, India}
\author{Z.~Ahammed}\affiliation{Variable Energy Cyclotron Centre, Kolkata 700064, India}
\author{J.~Amonett}\affiliation{Kent State University, Kent, Ohio 44242}
\author{B.D.~Anderson}\affiliation{Kent State University, Kent, Ohio 44242}
\author{D.~Arkhipkin}\affiliation{Particle Physics Laboratory (JINR), Dubna, Russia}
\author{G.S.~Averichev}\affiliation{Laboratory for High Energy (JINR), Dubna, Russia}
\author{S.K.~Badyal}\affiliation{University of Jammu, Jammu 180001, India}
\author{Y.~Bai}\affiliation{NIKHEF and Utrecht University, Amsterdam, The Netherlands}
\author{J.~Balewski}\affiliation{Indiana University, Bloomington, Indiana 47408}
% ...

\collaboration{STAR Collaboration}\noaffiliation

\date{\today}

\begin{abstract}
We present the centrality dependent measurement of multiplicity 
and pseudorapidity distributions of charged particles and photons 
in Au + Au collisions at $\sqrt{s_{\mathrm{NN}}}$ = 62.4 GeV.
The charged particles and photons are measured in the pseudorapidity 
region 2.9 $\le$ $\eta$  $\le$ 3.9 and 2.3 $\le$ $\eta$  $\le$ 3.7, 
respectively. We have studied the scaling of particle production 
with the number of participating nucleons and the number of 
binary collisions. The photon and charged particle production in 
the measured pseudorapidity range has been shown to be consistent 
with energy independent limiting fragmentation behavior. 
The photons are observed to follow a centrality independent
limiting fragmentation behavior while for the charged particles
it is centrality dependent. We have carried out 
a comparative study of the pseudorapidity distributions of positively 
charged hadrons, negatively charged hadrons, photons, pions, net 
protons in nucleus--nucleus collisions 
and pseudorapidity distributions from {\it p} + {\it p} collisions. From 
these comparisons we conclude that baryons in the 
inclusive charged particle distribution are responsible for the 
observed centrality dependence of limiting fragmentation. The mesons 
are found to follow an energy independent behavior of limiting 
fragmentation while the behavior of baryons seems to be energy dependent.
\end{abstract}

\maketitle

\section{INTRODUCTION}
The STAR experiment~\cite{star_nim} at the Relativistic 
Heavy Ion Collider (RHIC) at Brookhaven National 
Laboratory has the unique capability of measuring 
charged particle and photon multiplicities, precisely and 
simultaneously, at forward rapidity. By using this capability
we can carry out a systematic study of various aspects of 
charged particle and photon production in relativistic 
heavy ion collisions.

The conventional way of describing particle production in heavy
ion collisions is by measuring the particle density in pseudorapidity ($\eta$).
Within the framework of certain model assumptions, it
provides information on energy density, initial temperature and 
velocity of sound in the medium formed in the collisions~\cite{initial}. 
The widths of the pseudorapidity distributions are sensitive 
to longitudinal flow and re-scattering effects~\cite{ags,width}.
The variation of particle density in $\eta$ with collision centrality,
expressed in terms of the number of participating nucleons ($N_{\mathrm{part}}$)
and/or the number of binary collisions ($N_{\mathrm{coll}}$), can shed light on 
the relative importance of soft versus hard processes in particle production. 
The particle density in pseudorapidity also provides a test 
ground for various particle production models, 
such as those based on ideas of parton saturation~\cite{partonsaturation} and 
semi-classical QCD, also known as the color glass condensate (CGC)~\cite{cgc}.

At RHIC, the particle production mechanism could be different in
different regions of pseudorapidity. At midrapidity a significant increase 
in charged particle production normalized to the number of participating 
nucleons  has been observed from peripheral to 
central Au + Au collisions~\cite{phenix}. This has been 
attributed to the onset of hard  scattering processes, which scale 
with the number of binary collisions. However,
the total charged particle multiplicity per participant pair, integrated 
over the whole pseudorapidity range, is independent of 
centrality in Au + Au collisions~\cite{phobos}. In the framework of the 
color glass condensate picture of particle production~\cite{cgc}, 
the centrality dependence of particle production at midrapidity 
reflects the increase of gluon density due to the decrease in the 
effective strong coupling constant.
It will be interesting to see how the photon and charged particle 
production  scales with the number of participating nucleons and with the 
number of 
binary collisions in a common $\eta$ coverage at forward rapidity. 
The increase in particle production at midrapidity with increasing
center of mass energy has been studied in detail at RHIC~\cite{phobos}. 
It is also of interest to see how particle production varies with center of 
mass energy at forward rapidity. The experimental data on hadron 
multiplicity and its energy, centrality and 
rapidity dependence so far have been  
consistent with the approach based 
on ideas of parton saturation. Recently it has been argued that 
this onset of saturation occurs somewhere in the center of mass energy 
($\sqrt{s_{\mathrm{NN}}}$) range  of 17 GeV to 130 GeV~\cite{kharzeev1}. 
This is one 
of the reasons cited for having different mechanisms of particle production at 
RHIC and SPS. The present experimental data at $\sqrt{s_{\mathrm{NN}}}$ = 62.4
GeV may help to understand the transition 
energy for the onset of saturation effects in particle production.

It has been observed that inclusive photon production
(primarily from decay of $\pi^0$) at $\sqrt{s_{\mathrm{NN}}}$ =
62.4 GeV~\cite{starphoton} follows a centrality independent 
limiting fragmentation~\cite{limiting_frag} behavior. The inclusive 
charged particles at 19.6 GeV and 200 GeV have been observed to follow a 
centrality dependent behavior of limiting fragmentation~\cite{phobos}. 
It has been speculated that the baryons, an important 
constituent of inclusive charged particles, are responsible for 
the observed difference between photons and charged 
particles~\cite{phobos,starphoton}. 
The baryons coming from nuclear remnants and baryon transport, both of which 
change with centrality, may be the source of the centrality dependent 
limiting fragmentation for inclusive charged particles. The role of a new
mechanism of baryon production as discussed in Refs.~\cite{baryon_junction,valence_quark}
also needs to be understood. 
 A comparative study of
limiting fragmentation of positively and negatively charged particles and
photons at the same collision energy and pseudorapidity interval as
provided by the present data will help to understand the sources
responsible for the observed features.
On the theoretical side,
reproducing the energy, centrality, and species 
dependence of limiting fragmentation observed in the experimental data 
can be a good test for various particle production models. One such 
attempt to explain the energy dependence of limiting 
fragmentation phenomena within the framework of CGC has been carried out
in Ref.~\cite{LF_CGC}. The importance of the limiting fragmentation curve
on energy dependence of particle production has been 
demonstrated in Ref.~\cite{LF}.

 Event-by-event measurements of photon and charged particle multiplicities
 can be used to study multiplicity fluctuations~\cite{wa98_fluc}. 
 Fluctuations in physical 
 observables in heavy ion collisions may provide important 
 information regarding the formation of a Quark-Gluon Plasma and help to 
 address the question of thermalization~\cite{fluc_gen}.  The study of 
 event-by-event fluctuations in the ratio of photon to 
 charged particle multiplicities has also been proposed as a tool to search
 for production of Disoriented Chiral Condensates (DCCs)~\cite{dcc}.

In this paper we address some of the above physics issues through the 
first simultaneous measurement of the charged particle 
and photon multiplicities for Au + Au collisions 
at $\sqrt{s_{\mathrm{NN}}}$ = 62.4 GeV in the forward rapidity.
The charged particles are detected using the Forward Time 
Projection Chamber (FTPC) and the photons are detected using the 
Photon Multiplicity Detector (PMD) in the STAR 
experiment~\cite{star_nim,starftpc_nim,starpmd_nim}.

The paper is organized as follows: In section II we briefly describe 
the detectors used for measuring the charged particle and photon 
multiplicities and the trigger detectors used for selecting the minimum bias 
data, used in the present analysis. In section III we present the details of 
data analysis from the FTPC and the PMD. 
In section IV we present the results in 
terms of multiplicity and pseudorapidity distributions of 
photons and charged particles, scaling of particle production with number
of participating nucleons and number of binary collisions and limiting 
fragmentation behavior for charged, neutral and identified particles. 
Finally we summarize our study in section V.

\section{DETECTORS}
The STAR experiment~\cite{star_nim} consists of several detectors to measure 
hadronic and electromagnetic observables spanning
a large region of the available phase space at RHIC. 
The detectors used in the present analysis are the 
FTPC, PMD, a set of trigger detectors used for obtaining 
the minimum bias data and the Time Projection Chamber (TPC), 
data of which is used to determine 
the collision centrality. The FTPCs, PMD, minimum bias trigger
and collision centrality selection are briefly described below.

\subsection{Forward Time Projection Chambers}
There are two FTPCs; they are located on each side of the nominal 
collision vertex, around the beam axis.
They are cylindrical in structure with a diameter 
of 75 cm and 120 cm in length. Each FTPC has 10 rows of readout 
pads, called {$\mathrm {padrows}$},
which in turn are subdivided into 6 sectors with 160 pads per sector. 
The first padrow is located about 1.63 meters away on both sides 
from the center of the TPC (the nominal collision point). The sensitive medium 
is a gas mixture of Ar and CO$_2$ in the ratio of 50\%:50\% by weight.
The FTPCs detect charged particles in the pseudorapidity region 
2.5 $\le$ $\mid\eta\mid$ $\le$ 4.0. The novel design of the 
FTPCs uses a radial 
drift field, perpendicular to the magnetic field, to achieve a two-track 
resolution up to 2 mm. This allows for track reconstruction in the 
environment of high particle density at forward rapidity. In the present 
analysis, the data from only the FTPC in the positive pseudorapidity region 
(2.9 $\le$ $\eta$ $\le$ 3.9) is used. 
Particle production models 
such as HIJING~\cite{hijing} and AMPT~\cite{ampt} show that 
about 6--7\% of the total charged particles produced fall within 
the acceptance of each of the FTPCs. Further details of the design
characteristics of the FTPC can be found in Ref.~\cite{starftpc_nim}.

\subsection{Photon Multiplicity Detector}
The PMD is located 5.4 meters away from the center of the TPC 
(the nominal collision point) along the beam axis.
It consists of two planes (charged particle veto and preshower)
of an array of cellular gas proportional counters.
A lead plate of 3 radiation length thickness is placed between
the two planes and is used as a photon converter.
The sensitive medium is a gas mixture of Ar and
CO$_2$ in the ratio of 70\%:30\% by weight.
There are 41,472 cells in each plane, which are placed inside 
12 high voltage insulated and gas-tight chambers called supermodules. 
A photon traversing the converter produces an electromagnetic shower 
in the preshower plane, leading to a larger signal spread over several 
cells as compared to a charged particle, which is essentially confined 
to one cell. The PMD detects photons in the pseudorapidity region 
2.3 $\le$ $\eta$ $\le$ 3.7. In the present analysis, only the data 
from the preshower plane has been used. 
From HIJING~\cite{hijing} and AMPT~\cite{ampt} we find that
about 10--11\% of the total photons produced fall within 
the acceptance of the PMD. The photon production 
is dominated by photons from the decay of $\pi^{0}$s~\cite{starphoton}. 
HIJING calculations indicate that about 93--96\% of photons 
are from inclusive $\pi^0$ decays. Further details of the design
and characteristics of the PMD can be found in Ref.~\cite{starpmd_nim}. 

\subsection{Minimum bias trigger and collision centrality}
The minimum bias trigger is obtained using the charged particle hits
from an array of scintillator slats arranged in a barrel, called the 
Central Trigger Barrel, surrounding the TPC, 
two zero degree hadronic calorimeters at $\pm$18 m from the detector
center along the beam line, and two Beam-Beam Counters~\cite{trigger}. 
The centrality determination 
in this analysis uses the uncorrected multiplicity of charged particles in the 
pseudorapidity region $\mid\eta\mid$ $<$ 0.5, as measured by the 
TPC~\cite{star_glauber}. Table~\ref{table1} gives the percentage cross section, 
the corresponding uncorrected multiplicity of charged particle tracks 
($N_{\mathrm{ch}}^{\mathrm{TPC}}$) in the 
pseudorapidity region $\mid\eta\mid$ $<$ 0.5, the number of participating 
nucleons ($N_{\mathrm{part}}$) and the number of binary collisions 
($N_{\mathrm{coll}}$) used in this paper.
The number of participating nucleons and the number of binary collisions have been 
obtained from Glauber calculations~\cite{star_glauber}.

\begin{table}
\caption{ Centrality selection, number of participating nucleons and
number of binary collisions.
\label{table1}}
\begin{tabular}{ccccc}
\tableline
\% cross section&$N_{\mathrm{ch}}^{\mathrm{TPC}}$ &$\langle N_{\mathrm{part}} \rangle$
&$\langle N_{\mathrm{coll}} \rangle$\\
\tableline
0--5  &  $>$ 373   & 347.3  & 904.3\\
5--10 &  373--313  & 293.3  & 713.7\\
10--20 &  313--222 & 229.0  & 511.8\\
20--30 &  222--154 & 162.0  & 320.9\\
30--40 &  154--102 & 112.0  & 193.5\\
40--50 &  102--65  & 74.2   & 109.3\\
50--60 &  65--38   & 45.8   & 56.6 \\
60--70 &  38--20   & 25.9   & 26.8 \\
70--80 &  20--9    & 13.0   & 11.2\\
\tableline
\end{tabular}
\end{table}

\section{DATA RECONSTRUCTION}

\subsection{Charged particle reconstruction}
The analysis of the data from the FTPC involves the following steps:
(a) event selection, (b) pad-to-pad gain calibration, and (c) 
reconstruction of charged tracks.

A total of 1.2 million minimum bias events, 
corresponding to 0--80\% of the Au + Au hadronic interaction cross section, 
have been selected with a collision vertex position less than 30 cm 
from the center of the TPC along the beam axis.

The calibration of the FTPC is done using a laser 
calibration system~\cite{starftpc_nim}. This system helps to calibrate 
the drift velocity in the nonuniform radial drift field and also provides 
information for corrections to spatial distortions caused by mechanical 
or drift field imperfections. The localization of dead pads is done with
pulsers and by an analysis of data to identify electronically noisy 
pads~\cite{ftpcthesis}.

The reconstruction of experimental data 
involves two steps: (a) cluster-finding to calculate the track 
points from the charge distribution detected by the pads, (b) track-finding  
to group the track points of different padrows of the FTPC to form a
track. The cluster-finding includes reading of the
electronic signal data from the data acquisition system, looking for areas
of nonzero charge (cluster), deconvolution of clusters, and then 
finding the point co-ordinates. This is followed by combining 
clusters from all padrows to form tracks using a suitable tracking 
algorithm~\cite{markus}. A track is considered
valid if it consists of at least 5 found clusters and if its distance of
closest approach to the primary vertex is less than 3 cm. The 
condition of having at least 5 found clusters for each track in the 
FTPC ensures a small contribution of split tracks. The split tracks 
contribution and background contamination are primarily from $\gamma$ 
conversion electrons and positrons which are significantly reduced when 
we include those tracks which have transverse momentum in
the range 0.1 $<$ $p_{\mathrm{T}}$ $<$ 3 GeV/$c$ in the analysis. 
The maximum percentage of split tracks was estimated from 
simulations to be $\sim$1.5\%. 
The relative amount of split tracks decreases as we go from from central 
to peripheral collisions. 
Two procedures are used to obtain the charged particle 
yields at all $p_{\mathrm{T}}$.
The charged particle transverse 
momentum spectra are fitted by a power-law function in the range 
0.1 GeV/$c$ $<$ $p_{\mathrm{T}}$ $<$ 1 GeV/$c$ and extrapolated to 
$p_{\mathrm{T}}$ = 0 GeV/$c$. The 
low $p_{\mathrm{T}}$ yield is obtained from this extrapolation.
The other procedure calculates the yield of charged particles for 
$p_{\mathrm{T}}$ $<$ 0.1 GeV/$c$ by using the ratio of the 
yield in this $p_{\mathrm{T}}$ range to total 
yields from HIJING~\cite{hijing} simulations.
Both these procedures resulted in similar  
correction factors of the order of 15\% in the 
region 2.9 $\le$ $\eta$ $\le$ 3.9.

\begin{figure}
\begin{center}
\includegraphics[scale=0.35]{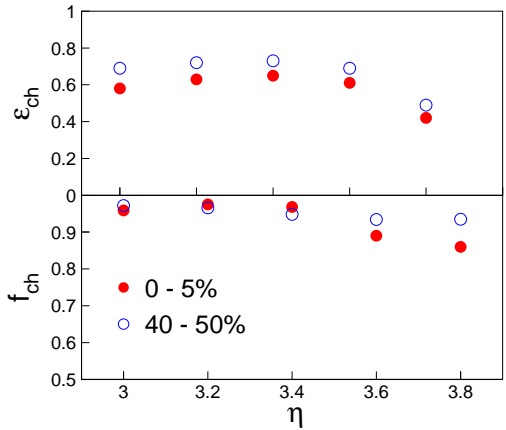}
\caption{(Color Online) Charged particle reconstruction efficiency ($\epsilon_{\mathrm{ch}}$)
and purity of charged hadron sample ($f_{\mathrm{ch}}$) in the 
FTPC as a function 
of pseudorapidity ($\eta$) for charged tracks with 
0.1 GeV/$c$ $<$ $p_{\mathrm{T}}$ $<$ 3 GeV/$c$, for two centrality intervals.}
\label{fig1}
\end{center}
\end{figure}
The efficiency of charged particle reconstruction ($\epsilon_{\mathrm{ch}}$) 
as a function of pseudorapidity is estimated by embedding Monte Carlo 
charged tracks into real data and then following the full 
reconstruction chain~\cite{starmultiplicityprl}. 
The reconstruction efficiency is obtained
by dividing the number of reconstructed Monte Carlo tracks within 
an $\eta$ bin by the total number of embedded Monte Carlo tracks 
in the same $\eta$ bin. The charged particle reconstruction efficiencies 
for central and peripheral collisions, for the $\eta$ region studied, 
are shown in Fig.~\ref{fig1}. The background 
contamination is obtained from detailed Monte Carlo 
simulation using the HIJING (version 1.382) 
event generator~\cite{hijing} and the 
detector simulation package  GEANT~\cite{geant}, which incorporates 
the full STAR detector framework.
The purity of the charged hadron sample ($f_{\mathrm{ch}}$) in the FTPC 
for central and peripheral collisions is also shown 
in Fig.~\ref{fig1}. The errors on efficiency and purity values will be
discussed later.

\subsection{Photon reconstruction}

The analysis of the data from the PMD involves the following steps:
(a) event selection, (b) cell-to-cell gain calibration, and (c) 
reconstruction or extraction of photon multiplicity. 

A total of 0.3 million minimum bias events, corresponding to 0--80\% of 
the Au + Au hadronic interaction  cross section, have been selected with 
a collision vertex position less than 30 cm from the center of the 
TPC along the beam axis. The difference in the number of events for the PMD
and FTPC analysis originates from the fact that, for the same period 
of data-taking, the PMD 
recorded fewer events and there was a need for a more stringent 
data clean-up procedure to remove events with pile-up-like effects.

The cell-to-cell gain calibration was done by obtaining the ADC distributions 
of isolated cells. The ADC distribution of an isolated cell may be treated 
as the response of the cell to charged particles~\cite{starpmd_nim}.
For most of the cells this response follows a Landau distribution. 
We use the mean of the ADC distribution of isolated cells
to estimate and correct the relative gains of all cells within each
supermodule. The cell-to-cell gain variation is between 10--25\%
for different supermodules.

The extraction of the photon multiplicity proceeds in two steps involving  
clustering of hits and photon-hadron discrimination. Hit clusters consist of 
contiguous cell signals. Photons are separated from charged particles using the
following conditions: (a) the number of cells in a cluster is  $>$ 1, and (b) 
the cluster signal is larger than 3 times the average response 
of all isolated cells in a supermodule. The choice of the conditions 
is based on a detailed study of simulations~\cite{starpmd_nim,starphoton}. 
The number of selected clusters, called
$\gamma-like$  clusters ($N_{\mathrm {\gamma-like}}$), in different
supermodules for the same $\eta$ coverage are used to evaluate 
the effect of possible
non-uniformity in the response of the detector. 

\begin{figure}
\begin{center}
\includegraphics[scale=0.35]{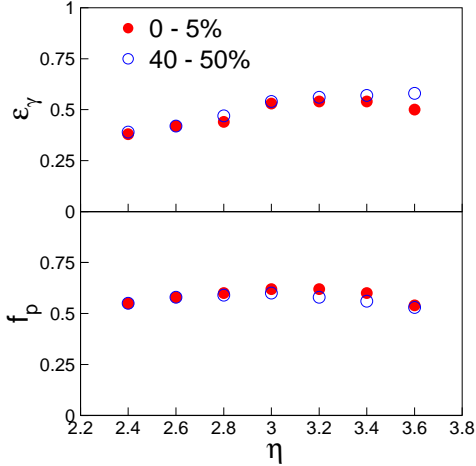}
\caption{(Color Online) Photon reconstruction efficiency ($\epsilon_{\mathrm {\gamma}}$) and 
purity of photon sample ($f_{\mathrm {p}}$) 
for PMD as a function of pseudorapidity ($\eta$), for two centrality intervals.
}
\label{fig2}
\end{center}
\end{figure}

To estimate the number of photons ($N_{\mathrm {\gamma}}$) from the detected 
$N_{\mathrm {\gamma-like}}$ clusters we evaluate the photon reconstruction
efficiency ($\epsilon_{\mathrm {\gamma}}$) and purity ($f_{\mathrm {p}}$) of 
the $\gamma-like$ sample defined~\cite{wa98_dndy} 
as $\epsilon_\gamma  =  N^{\gamma,th} _{cls} / N_\gamma$
and $f_{\mathrm {p}}              =  N^{\gamma,th} _{cls} / N_{\gamma-like}$ respectively.
$N_{cls}^{\mathrm {\gamma,th}}$ is the number of photon clusters above the 
photon-hadron discrimination condition. Both $\epsilon_{\gamma}$ 
and $f_{\mathrm{p}}$ are obtained from a detailed Monte Carlo simulation using 
HIJING~\cite{hijing} 
with default parameter settings and the detector simulation package 
GEANT~\cite{geant}, which incorporates 
the full STAR detector framework.
Both $\epsilon_{\gamma}$ and $f_{\mathrm{p}}$ 
vary with pseudorapidity and centrality. This is
due to variations in particle density, upstream conversions  and detector
related effects. A photon should ideally create one cluster in 
the detector. However, it may 
give rise to more than one cluster (called split clusters) in the real 
experimental environment. These may happen because of conversions of the 
photon due to upstream materials in front of the PMD, or limitations of the 
clustering algorithm due to varying particle density. The highest 
occupancy of the PMD is about 12\% and the maximum percentage of 
split clusters is estimated to be 9\%.
The photon reconstruction efficiency and the purity of the photon sample 
determined by means of simulations for central and peripheral collisions for
the $\eta$ region studied are shown in 
Fig.~\ref{fig2}.
The lower limit of photon $p_{\mathrm{T}}$ acceptance in the PMD is estimated 
from detector simulations to be 20 MeV/$c$. 

\subsection{Systematic errors}
The systematic errors on the charged particle multiplicity 
($N_{\mathrm{ch}}$) are due to uncertainties in estimates of 
$\epsilon_{\mathrm{ch}}$ and $f_{\mathrm{ch}}$. The uncertainty in the estimates
are obtained through simulations by varying the track quality cuts. 
The value of the maximal distance of closest approach of a track to the 
primary vertex is 
varied by 0.5 cm leading to a maximum error on $N_{\mathrm{ch}}$ of
$\sim$6\%. The minimum number of clusters to form a track was varied 
from 5 to 4. This led to an error on $N_{\mathrm{ch}}$ of $\sim$1\%. The
uncertainty in the correction factor to obtain the $N_{\mathrm{ch}}$ yield for 
$p_{\mathrm{T}}$ $<$ 0.1 GeV/$c$ is $\sim$8\%. This also contributes to the total 
systematic errors. The total systematic error in $N_{\mathrm{ch}}$ is 
$\sim$10\% for all the centrality classes studied. The systematic error
for the region $\eta$ $>$ 3.6 is estimated to be about 15\%, due to 
larger uncertainity in the reconstruction efficiency. This arises
primarily due to uncertainity in realistic reproduction of electronic loss, 
at the extreme ends of the detector acceptance. This is estimated by studying the
azimuthal dependence of charged particle density in a given $\eta$ window.

The systematic errors on the photon multiplicity ($N_{\mathrm{\gamma}}$) 
are  due to 
(a) uncertainty in estimates of  $\epsilon_{\mathrm {\gamma}}$ and 
$f_{\mathrm {p}}$  values 
    arising from splitting of clusters and the choice of 
    photon-hadron discrimination threshold and  
(b) uncertainty in $N_{\mathrm{\gamma}}$  arising from the 
non-uniformity of the detector response primarily due 
to cell-to-cell gain variation. 
The error in $N_{\mathrm{\gamma}}$ due to (a) is
estimated from Monte Carlo simulations to be 9.8\% and 7.7\% in 
central and peripheral collisions, respectively.
The error in $N_{\mathrm{\gamma}}$ due to (b) is
estimated using average gains for normalization and by studying the 
azimuthal dependence of the photon density of the detector in an $\eta$ window
to be 13.5\% for central and 15\% for peripheral collisions. 
The total systematic error in $N_{\mathrm{\gamma}}$ is $\sim$17\% 
for both central and peripheral collisions. 

The total errors on $N_{\mathrm{ch}}$ and $N_{\mathrm{\gamma}}$ are obtained 
by adding respective systematic and statistical errors in quadrature and are 
shown in all the figures unless mentioned otherwise. The statistical 
errors are small and within the symbol sizes.

\section{RESULTS and Discussion}

\subsection{Multiplicity distributions}
\begin{figure}
\begin{center}
\includegraphics[scale=0.4]{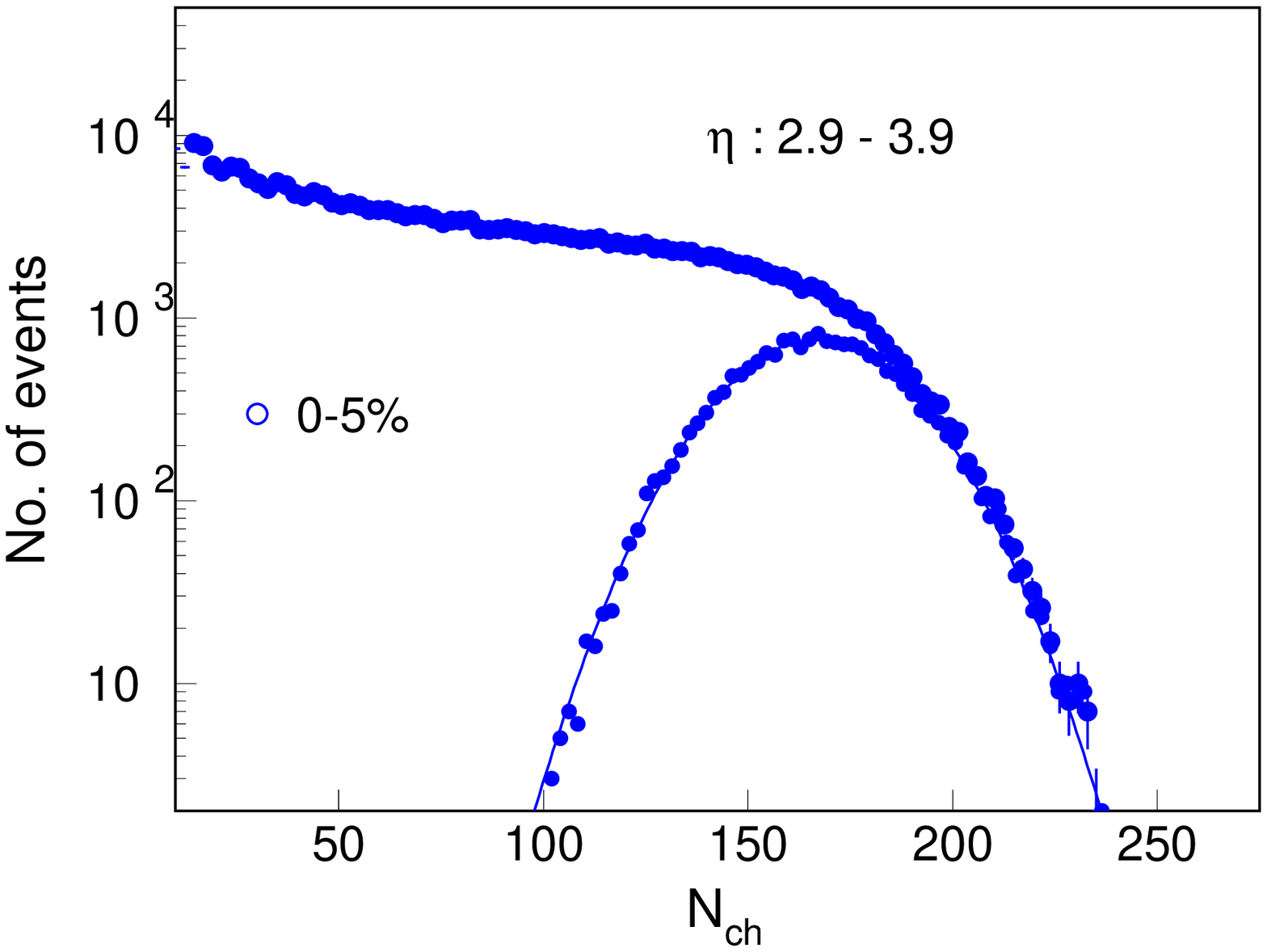}
\includegraphics[scale=0.4]{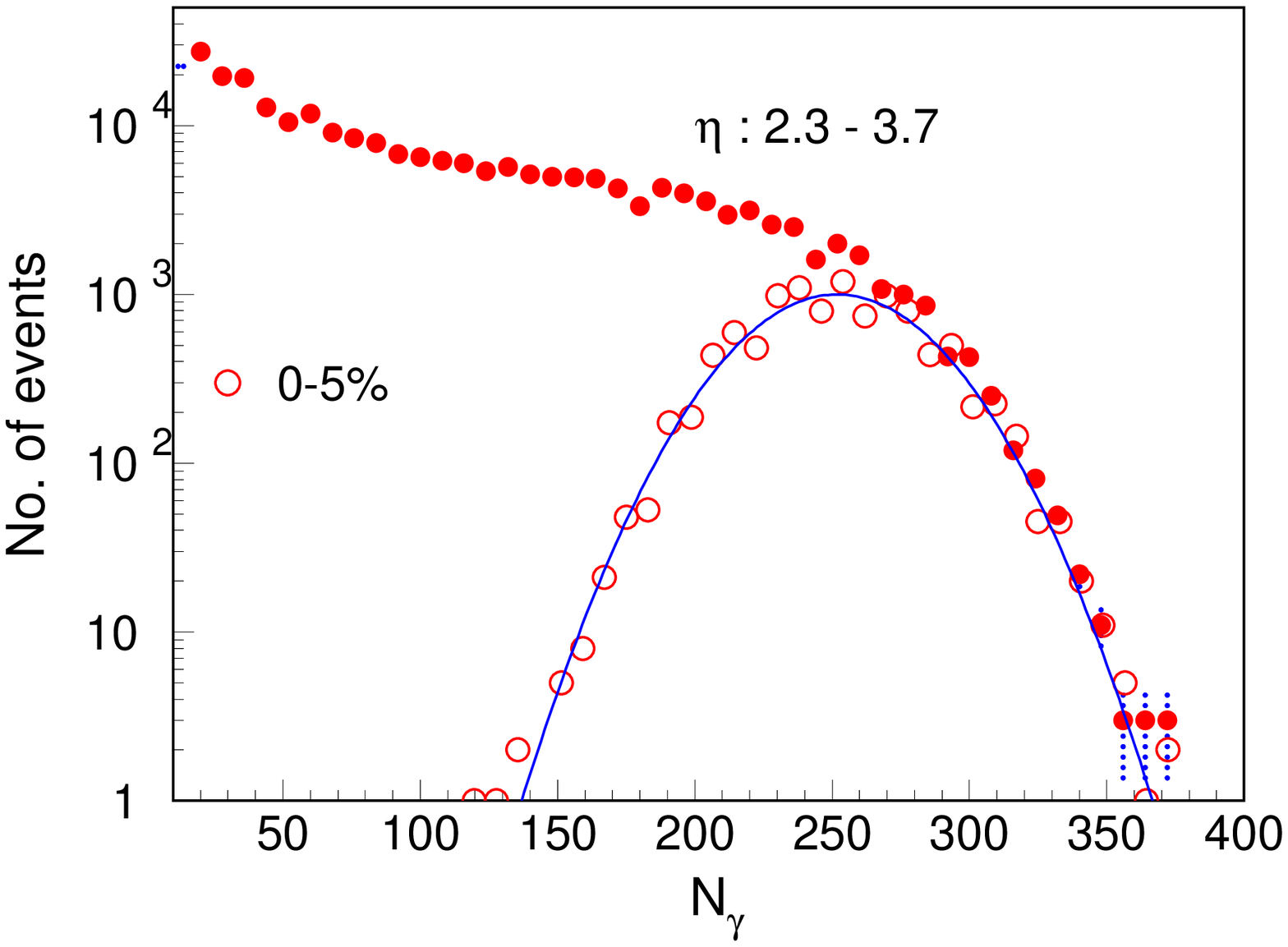}
\caption{ (Color Online) Minimum bias $N_{\mathrm{ch}}$  
(2.9 $\le$ $\eta$  $\le$ 3.9 ) and $N_{\mathrm{\gamma}}$  
(2.3 $\le$ $\eta$  $\le$ 3.7) distributions for Au + Au collisions at
$\sqrt{s_{\mathrm{NN}}}$ = 62.4 GeV. The charged particle and 
photon multiplicity 
distribution for top 5\% central events are shown in open circles.
The solid curve is the Gaussian fit to the data points.}
\label{fig3}
\end{center}
\end{figure}
\begin{figure}
\begin{center}
\includegraphics[scale=0.4]{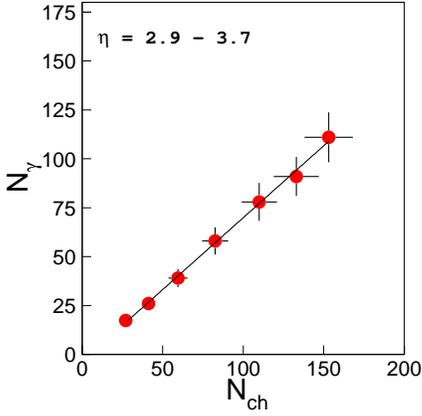}
\caption{(Color Online) Correlation between average number
of charged particles ($N_{\mathrm{ch}}$) 
and average number of photons ($N_{\mathrm{\gamma}}$) within 
the common $\eta$ range of FTPC and PMD 2.9 $\le$ $\eta$  $\le$ 3.7 
for different collision centrality classes
in Au + Au collisions at $\sqrt{s_{\mathrm{NN}}}$ = 62.4 GeV. The solid line is a 
straight line fit to the data points (see text for details).}
\label{fig3a}
\end{center}
\end{figure}

The charged particle multiplicity ($N_{\mathrm{ch}}$) 
and photon multiplicity ($N_{\mathrm{\gamma}}$) are obtained event-by-event
in the FTPC and the PMD following the analysis procedure described above. 
Fig.~\ref{fig3} shows the minimum bias distributions of
$N_{\mathrm{ch}}$ and $N_{\mathrm{\gamma}}$ for Au + Au collisions at 
$\sqrt{s_{\mathrm{NN}}}$ = 62.4 GeV.
The distributions have a characteristic
shape with a steep rise that corresponds to the most
peripheral events. The plateaus in the photon and charged particle
multiplicity distributions correspond to mid-central events and the
fall-off to the most central collision events. The shape of
the curves in the fall-off region reflects 
the intrinsic fluctuations of the measured quantities and
the limited acceptance of the detectors.
The event-by-event 
charged particle and photon multiplicity distributions for 0--5\% 
central collisions are also shown. 
Gaussian fits to these distributions have been made. The values of the fit 
parameters for charged particles measured in 2.9 $\le$ $\eta$  $\le$ 3.9 are: 
mean = 167 and $\sigma$ = 20; $\chi^{2}/ndf$ = 70.67/69. The values of the
fit parameters for photons measured in  2.3 $\le$ $\eta$  $\le$ 3.7 are: 
mean = 252 and $\sigma$ = 30; $\chi^{2}/ndf$ = 37.3/34. 
The correlation between the 
average number of charged particles and average number of photons within 
the common pseudorapidity coverage of the FTPC and PMD 
(2.9 $\le$ $\eta$  $\le$ 3.7) for 
different collision centrality classes in Au + Au collisions at 
$\sqrt{s_{\mathrm{NN}}}$ = 62.4 GeV are shown in Fig.~\ref{fig3a}. 
The correlation between $N_{\mathrm{ch}}$ and $N_{\mathrm{\gamma}}$
can be expressed as $N_{\mathrm \gamma}$ = (0.74$\pm$0.01)$N_{\mathrm{ch}}$
-- (3.57$\pm$0.83). This is shown as a straight
line in the figure. The correlation reflects the variation of 
$N_{\mathrm{\gamma}}$ and $N_{\mathrm{ch}}$ with collision centrality.
The correlation coefficient is 1.01$\pm$0.01.

\subsection{Scaling of particle production}

\begin{figure}
\begin{center}
\includegraphics[scale=0.45]{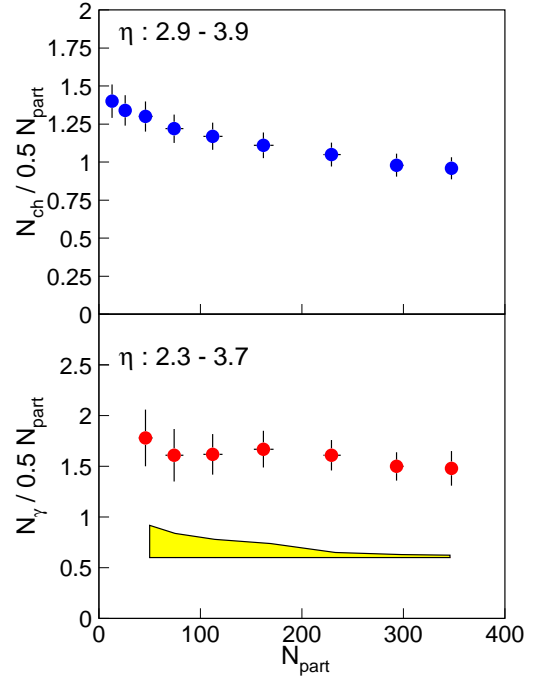}
\caption{ (Color Online) Variation of $N_{\mathrm{ch}}$ normalized to the 
number of participating  nucleon
pair in the FTPC 
coverage (2.9 $\le$ $\eta$  $\le$ 3.9) and $N_{\mathrm{\gamma}}$
normalized to the number of participating nucleon pair in the 
PMD acceptance (2.3 $\le$ $\eta$  $\le$ 3.7) 
as a function of $N_{\mathrm{part}}$. The lower band shows the
uncertainty in the ratio due to uncertainties in 
$N_{\mathrm{part}}$ calculations. 
}
\label{fig7}
\end{center}
\end{figure}

After having discussed the event-by-event measurement of photon and 
charged particle multiplicities in the previous section, we now discuss 
the variation of average (averaged over number of events) 
photon and charged particle multiplicities
within the full coverage of the PMD and FTPC, respectively, with centrality. 
Collision centrality is expressed in terms of either number
of participating nucleons or number of binary collisions. This 
will provide information on the contribution of hard (pQCD jets) 
and soft processes to particle production at forward rapidity. 
The scaling of particle production with the number of participating 
nucleons indicates the dominance of soft processes 
while scaling with the number of binary collisions indicates the onset of
hard processes. At midrapidity the particle production at $\sqrt{s_{\mathrm{NN}}}$ =
130 GeV and 200 GeV has been shown to scale with a combination of
$N_{\mathrm{part}}$ and $N_{\mathrm{coll}}$~\cite{phenix}.
Here we present the results on scaling of particle production at
forward rapidity for Au + Au collisions at $\sqrt{s_{\mathrm{NN}}}$ = 62.4 GeV. 

Figure~\ref{fig7} shows the variation of the total 
number of charged particles in the FTPC coverage 
(2.9 $\le$ $\eta$  $\le$ 3.9) and the
total number of photons in the PMD acceptance
(2.3 $\le$ $\eta$  $\le$ 3.7), both normalized to $N_{\mathrm{part}}$,
 as a function of the collision centrality, expressed by 
the number of participants. 
Higher $N_{\mathrm{part}}$ values correspond to more central collisions, or 
collisions with smaller impact parameter. The charged particle 
yield per participating nucleon pair at forward rapidity decreases from 
peripheral to central collisions. The photon production per participant pair 
is found to be approximately constant with centrality in the 
forward $\eta$ range studied. 
\begin{figure}
\begin{center}
\includegraphics[scale=0.45]{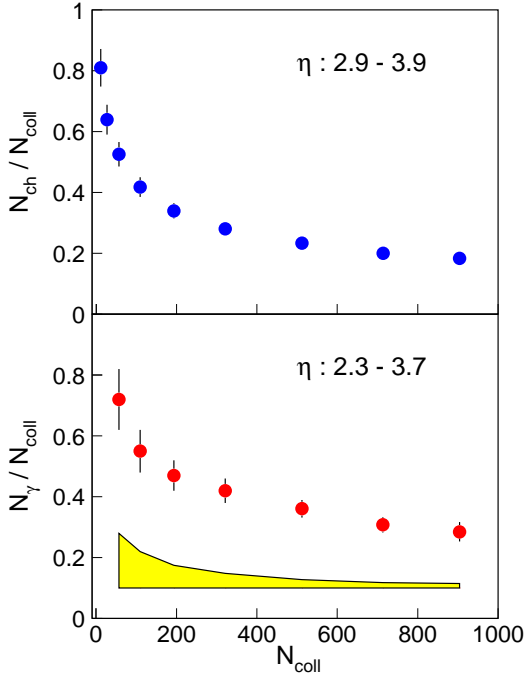}
\caption{ (Color Online) Variation of $N_{\mathrm{ch}}$ normalized to
the number of collisions in the FTPC 
coverage (2.9 $\le$ $\eta$  $\le$ 3.9) and $N_{\mathrm{\gamma}}$
normalized to number of collisions, in the 
PMD coverage (2.3 $\le$ $\eta$  $\le$ 3.7) 
as a function of $N_{\mathrm{coll}}$. The lower band 
shows the uncertainty in the ratio due to uncertainties in
$N_{\mathrm{coll}}$ calculations. 
}
\label{fig8}
\end{center}
\end{figure}

Figure~\ref{fig8} shows the variation of the total number of charged particles 
normalized to the
number of collisions in the FTPC coverage (2.9 $\le$ $\eta$  $\le$ 3.9) and 
the total number of photons normalized to the number of collisions
 in the PMD coverage 
(2.3 $\le$ $\eta$  $\le$ 3.7) as a function of the number 
of binary collisions. 
Higher $N_{\mathrm{coll}}$ values correspond to more central collisions, or 
collisions with smaller impact parameter. Both the charged 
particle yield and photon yield normalized to the number of binary collisions 
do not scale with the number of binary collisions at forward rapidity. 
The data value decreases from peripheral to central collisions.
This indicates that the contribution of hard processes to particle production
at forward rapidity is small.

\subsection{Pseudorapidity distributions}
\begin{figure}
\begin{center}
\includegraphics[scale=0.45]{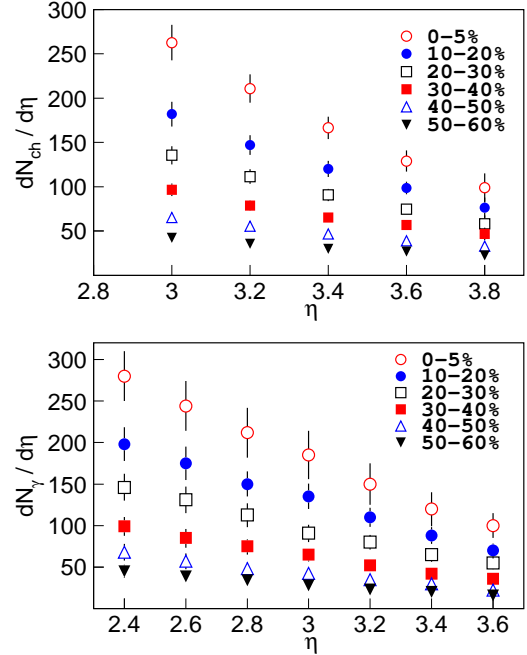}
\caption{ (Color Online) $dN/d\eta$ for charged particles 
and photons for
Au + Au collisions at $\sqrt{s_{\mathrm{NN}}}$ = 62.4 GeV  for various event 
centrality classes.
}
\label{fig4}
\end{center}
\end{figure}
So far we have discussed the multiplicities of photons and charged particles
over the full coverage of the detectors. In this section we 
study the variation in particle density with $\eta$.
The results can then be directly compared to different models 
in order to understand the mechanism of particle production in 
heavy ion collisions at forward rapidity. 

Figure~\ref{fig4} shows the pseudorapidity distributions of charged particles 
within 2.9 $\le$ $\eta$ $\le$ 3.9 and photons within 2.3 $\le$ $\eta$ $\le$ 3.7 for 
various event centrality classes. As expected the particle density increases with
decrease in $\eta$.
\begin{figure}
\begin{center}
\includegraphics[scale=0.45]{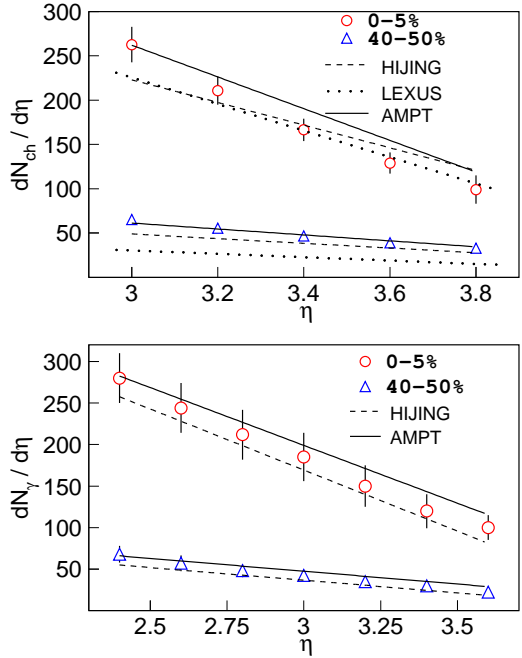}
\caption{ (Color Online) $dN/d\eta$ for charged particles 
and photons for 
central and peripheral Au + Au collisions at $\sqrt{s_{\mathrm{NN}}}$ = 62.4 GeV
compared to corresponding results from theoretical models.}
\label{fig5}
\end{center}
\end{figure}
Fig.~\ref{fig5} shows the comparison of pseudorapidity distributions
for photons and charged particles for 0--5\% and 40--50\% central
Au + Au collisions at $\sqrt{s_{\mathrm{NN}}}$ = 62.4 GeV with the corresponding 
results from various theoretical models.
The HIJING model~\cite{hijing} is based on 
perturbative QCD processes which lead to multiple jet production and 
jet interactions in matter. HIJING seems to 
underpredict the measured photon multiplicity. However within the 
systematic errors it is difficult to make definitive conclusions.
For charged particles, HIJING fails to explain the 
$\eta$ distributions for central and peripheral collisions.
The AMPT~\cite{ampt} model is a multi-phase transport model 
which includes both initial partonic and final hadronic interactions. 
For photons, the results from the AMPT model 
are in reasonable agreement with the data 
for central and peripheral events within the systematic errors. 
For charged particles in central collisions, the results from
AMPT explain the data at lower $\eta$ and overpredict the 
charged particle yields at higher $\eta$.
The LEXUS~\cite{lexus} model is based on linear extrapolation of
nucleon-nucleon collisions to high-energy 
nucleus-nucleus collisions.
For charged particles, the LEXUS model underpredicts the multiplicity
at lower $\eta$ and agrees with experimental data at higher $\eta$ 
for central collisions. It also underpredicts the charged particle yields
for peripheral collisions.
In summary, we observe that the photon multiplicity within the systematic errors
is reasonably well explained by HIJING and AMPT models. The detailed
pseudorapidity dependence of the charged particle multiplicity is not
reproduced by the above models.

\begin{figure}
\begin{center}
\includegraphics[scale=0.42]{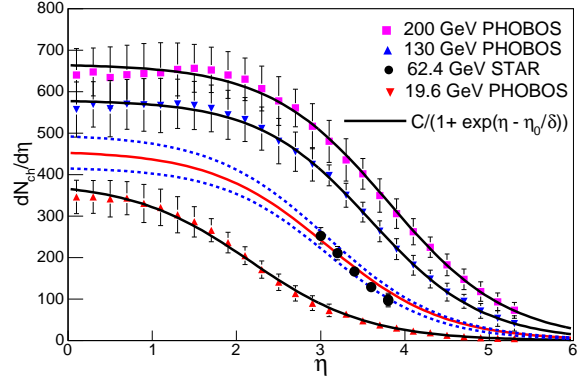}
\caption{ (Color Online) Pseudorapidity distributions of 
charged particles for various center of mass energies 
in Au + Au central collisions. The pseudorapidity
distributions for $\sqrt{s_{\mathrm {NN}}}$ = 200 GeV, 130 GeV and 19.6 GeV 
are from the PHOBOS experiment~\cite{phobos}.
The solid lines are the results of the fits
described in the text.}
\label{fig5a}
\end{center}
\end{figure}
We have so far studied the pseudorapidity distribution of particles
at forward rapidity at $\sqrt{s_{\mathrm {NN}}}$ = 62.4 GeV. 
Now we will investigate the following: (a) the energy dependence of 
the shape of the $\eta$ distribution of charged particles
available at various energies of Au + Au collisions in RHIC, and (b)
try to estimate the full $\eta$ distribution for charged particles
for $\sqrt{s_{\mathrm {NN}}}$ = 62.4 GeV from the above study 
and compare to the present measurements.

The full pseudorapidity distribution of charged particles at RHIC for central collisions
can be parametrized by the following 3-parameter formula:

\begin{equation}
\nonumber
\frac{dN}{d\eta} = \frac{C}{1 + {\exp} \frac{\eta - \eta_{0}}{\delta}}
\end{equation}

This formula is chosen to describe the central plateau and the fall
off in the fragmentation region of the distribution by means of the 
parameters $\eta_{0}$ and $\delta$ respectively. Using this formula
we can describe the 200 GeV, 130 GeV and 19.6 GeV pseudorapidity 
distributions of charged particles from the PHOBOS experiment~\cite{phobos}. The 
values of the parameters $C$,
$\eta_{0}$ and $\delta$ are given in Table~\ref{table2} and the fits to data 
are shown in Fig.~\ref{fig5a}. The value of $\eta_{0}$ is found to 
increase with increasing $\sqrt{s_{\mathrm{NN}}}$. The value of the parameter
$\delta$ is found to be independent of energy within errors.
This feature is another way of testing the concept of limiting
fragmentation, which will be discussed later. Using the average
value of $\delta$ and interpolating the value of $\eta_{0}$ to
62.4 GeV we are able to predict the full pseudorapidity distribution
for charged particles at 62.4 GeV. This is shown as solid curve 
in Fig.~\ref{fig5a}, together with our measured charged particle 
data for 62.4 GeV at forward rapidity. The dashed curves represent the 
error in obtaining the full pseudorapidity distribution for charged particles
using the interpolation method described.

\begin{table}
\caption{ Parameters $C$, $\eta_{0}$ and $\delta$ for different $\sqrt{s_{\mathrm {NN}}}$.
\label{table2}}
\begin{tabular}{cccc}
\tableline
$\sqrt{s_{\mathrm{NN}}}$ (GeV) & $C$ & $\eta_{0}$ & $\delta$\\
\tableline
19.6  &  382 $\pm$ 33 & 2.16 $\pm$ 0.17     & 0.7 $\pm$ 0.06  \\
62.4 (interpolated) & 458 $\pm$ 40  & 3.08 $\pm$ 0.35  & 0.69 $\pm$ 0.06  \\
130   &  580 $\pm$ 21 & 3.59 $\pm$ 0.076    & 0.66 $\pm$ 0.05  \\
200   &  667 $\pm$ 22 & 3.80 $\pm$ 0.082    & 0.71 $\pm$ 0.06  \\
\tableline
\end{tabular}
\end{table}

\begin{figure}
\begin{center}
\includegraphics[scale=0.45]{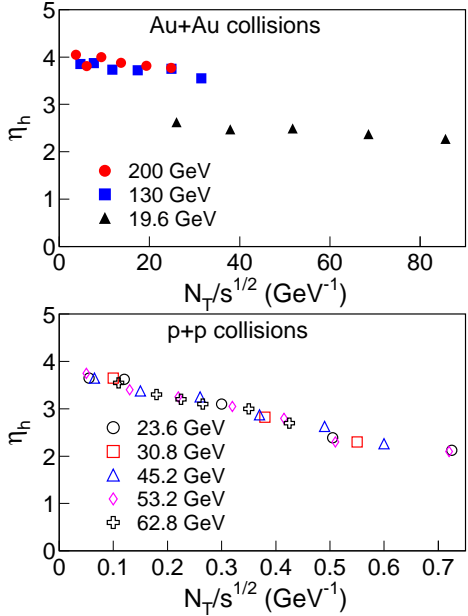}
\caption{ (Color Online) Half width at half maximum of the pseudorapidity
distributions ($\eta_{\mathrm h}$) of charged particles 
as a function of total charged particle multiplicity ($N_{\mathrm{T}}$) normalized 
to the center of mass energy. The Au + Au collision data are 
from the PHOBOS~\cite{phobos} experiment and {\it p} + {\it p} collision
data are from the ISR~\cite{isr} experiments.
}
\label{fig5b}
\end{center}
\end{figure}
We have also studied the widths of the pseudorapidity distributions
of charged particles at RHIC and compared them to those from {\it p} + {\it p} 
collisions
at ISR~\cite{isr}.
In Fig.~\ref{fig5b} we show the variation of the half width at half
maximum ($\eta_{\mathrm h}$) of the charged particle 
pseudorapidity distributions 
as a function of total charged particle multiplicity normalized
to the center of mass energy ($N_{\mathrm{T}}$/$\sqrt{s_{\mathrm{NN}}}$) 
for {\it p} + {\it p} and Au + Au collisions.
The data shown is for various centrality classes in Au + Au collisions~\cite{phobos} 
and for various intervals of observed total multiplicity in
{\it p} + {\it p} collisions. We observe that the half width at half-maximum 
obeys an interesting scaling law in {\it p} + {\it p} collisions and is found
to depend on a single variable ($N_{\mathrm{T}}$/$\sqrt{s_{\mathrm{NN}}}$).
In Au + Au collisions this scaling seems to be valid for 
200 GeV and 130 GeV. Although the width decreases with 
$N_{\mathrm{T}}$/$\sqrt{s_{\mathrm{NN}}}$ for 19.6 GeV, the data lies below
the higher energy data unlike the energy independent behavior 
observed in {\it p} + {\it p} collisions. This may reflect the change in the mechanism
of particle production over the full pseudorapidity range as we increase
the $\sqrt{s_{\mathrm{NN}}}$ from 19.6 GeV to $\sqrt{s_{\mathrm{NN}}}$ $>$ 130 GeV in Au + Au collisions at RHIC.

\subsection{Energy dependence of particle multiplicity}
\begin{figure}
\begin{center}
\includegraphics[scale=0.4]{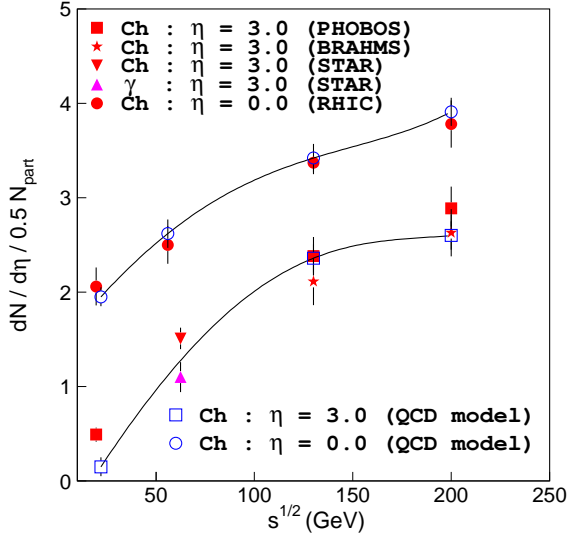}
\caption{ (Color Online) $dN/d\eta$ per participating nucleon pair at 
midrapidity ($\eta$ = 0) and forward rapidity ($\eta$ = 3.0) for 
various center of mass energies for central collisions. The data 
for charged particles at $\sqrt{s_{\mathrm{NN}}}$ = 19.6 GeV, 56 GeV, 130 GeV 
and 200 GeV 
at $\eta$ = 3.0 are from the PHOBOS~\cite{phobos} and BRAHMS~\cite{brahms} 
experiments. 
The data for charged particles at
midrapidity are the averages of the values from the 4 RHIC experiments. The photon 
yield at $\sqrt{s_{\mathrm{NN}}}$ = 62.4 GeV is also plotted. For comparison the results
from a model based on  parton saturation expected at high-density
QCD~\cite{kharzeev} are also shown. The solid lines are polynomial fits to the values from the 
QCD model. There is no prediction 
for $\sqrt{s_{\mathrm{NN}}}$ = 62.4 GeV available from this model. }
\label{fig6}
\end{center}
\end{figure}
The energy dependence of charged particle yields at midrapidity has been
studied at RHIC~\cite{phobos}. Here we present the results on the energy 
dependence of particle yields at forward rapidity and compare them 
with yields at midrapidity.

Figure~\ref{fig6} shows the charged particle pseudorapidity distribution
scaled by the number of
participating nucleon pairs at midrapidity ($\eta$ = 0) and forward rapidity
($\eta$ = 3.0) as a function of $\sqrt{s_{\mathrm{NN}}}$ for central collisions at 
RHIC. The data for charged particles at $\sqrt{s_{\mathrm{NN}}}$ = 19.6 GeV, 
56 GeV, 130 GeV 
and 200 GeV at $\eta$ = 3.0 are from the PHOBOS~\cite{phobos} and 
BRAHMS~\cite{brahms} experiments. 
The data for charged particles at midrapidity are the averages of the values 
from the 4 RHIC experiments. The charged particle production at $\eta$ = 0,
can be expressed as 
\begin{eqnarray}
\nonumber 
\frac{dN/d\eta}{0.5N_{\mathrm{part}}} = 
1.75 (\pm0.25) + 
0.017 (\pm0.005) \ln\left[\sqrt{s_{\mathrm{NN}}}\,\right] \nonumber\\
- 0.00003 (\pm0.00002) (\ln\left[\sqrt{s_{\mathrm{NN}}}\, \right])^{2} \nonumber.
\end{eqnarray}
The charged particle production at $\eta$ = 3.0,
can be expressed as 
\begin{eqnarray}
\nonumber 
\frac{dN/d\eta}{0.5N_{\mathrm{part}}} = 
-0.03 (\pm0.13) + 0.028 (\pm0.004) \ln\left[\sqrt{s_{\mathrm{NN}}}\,\right] \nonumber\\
-0.00007 (\pm0.00002) (\ln\left[\sqrt{s_{\mathrm{NN}}}\,\right])^{2}\nonumber.
\end{eqnarray}
The ratio of charged particle production 
at $\eta$ = 0 to that at $\eta$ = 3.0 decreases from a factor 
4 to 1.3 as $\sqrt{s_{\mathrm{NN}}}$
increases from 19.6 GeV to 200 GeV. 
The photon result at $\sqrt{s_{\mathrm{NN}}}$ = 62.4 GeV for $\eta$ = 3.0 
is also shown. The photon yields at other $\sqrt{s_{\mathrm{NN}}}$ values at forward 
rapidity and midrapidity are not yet available at RHIC. 
The photon production at 
$\sqrt{s_{\mathrm{NN}}}$ = 62.4 GeV is about 35\% lower than the 
charged particle 
production for the same energy at $\eta$ = 3.0. The charged 
particle yield at $\eta$ = 3.0 for $\sqrt{s_{\mathrm{NN}}}$ = 62.4 GeV is a 
factor 1.6 and 1.9 lower compared to the corresponding yields at 
130 GeV and 200 GeV and a factor 3.0 higher than the charged particle yields at 
19.6 GeV. 
For comparison, also shown in Fig.~\ref{fig6} are the results from a
model based on  parton saturation, which is 
expected in high-density QCD~\cite{kharzeev}. 
The results from the model agree with the measured charged particle 
yields at midrapidity for all energies at RHIC. However, the model's
prediction for forward rapidity at the lowest energy (22 GeV) is lower 
compared to data (19.6 GeV). 
There is no prediction for $\sqrt{s_{\mathrm{NN}}}$ = 62.4 GeV available 
from this model. It would be interesting to have the predictions
to understand the transition energy for the onset of saturation effects
at RHIC.   

\subsection{Comparison of $N_{\mathrm{ch}}$ and $N_{\mathrm{\gamma}}$ }
\begin{figure}
\begin{center}
\includegraphics[scale=0.4]{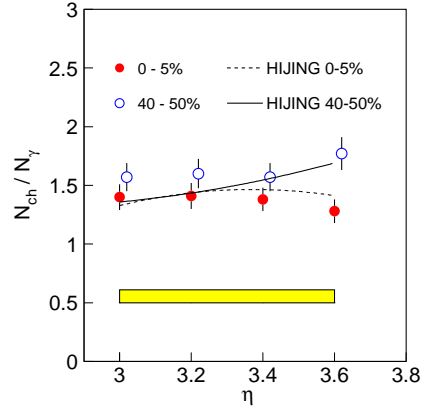}
\caption{(Color Online) Ratio of $N_{\mathrm{ch}}$ to $N_{\mathrm{\gamma}}$ 
for 0--5\%  and 40--50\% central Au + Au collisions
at  $\sqrt{s_{\mathrm{NN}}}$ = 62.4 GeV as a function of $\eta$.
Results from HIJING are also shown for comparison. The lower band reflects the
common errors in ratio for the two centrality classes.}
\label{fig61}
\end{center}
\end{figure}
The STAR experiment at RHIC has the unique capability to study the 
yields of charged particle and photons at forward rapidity.
Fig.~\ref{fig61} shows the ratio of $N_{\mathrm{ch}}$ to $N_{\mathrm{\gamma}}$
for 0--5\% and 40--50\% central Au + Au collisions
at  $\sqrt{s_{\mathrm{NN}}}$ = 62.4 GeV as a function of $\eta$ in the
common $\eta$ coverage of the FTPC and the PMD. 
The ratio is around 1.4 for central collisions and 1.6 for 
peripheral collisions within 3.0 $<$ $\eta$ $<$ 3.6. The results
from HIJING indicate similar values. The correlated systematic
errors, mainly arising due to uncertainties in the Monte Carlo
determination of reconstruction efficiencies and normalization errors,
are not plotted on the data points and are shown as a shaded band. 
The photon production
is dominated by photons from the decay of $\pi^{0}$s~\cite{starphoton}. 
The charged particle yields have a substantial contribution from baryons 
at forward rapidity~\cite{brahms_proton}. Apart from 
the kinematics, this may be the 
reason for higher charged particle yields compared to photons.
In the future, event--by--event study of $N_{\mathrm{ch}}$ and 
$N_{\mathrm{\gamma}}$
correlations in common $\eta$ and $\phi$ coverage of 
the FTPC and the PMD
can be used to look for possible formation of disoriented chiral 
condensates~\cite{dcc}.

\subsection{Energy dependence of limiting fragmentation}
\begin{figure}
\begin{center}
\includegraphics[scale=0.48]{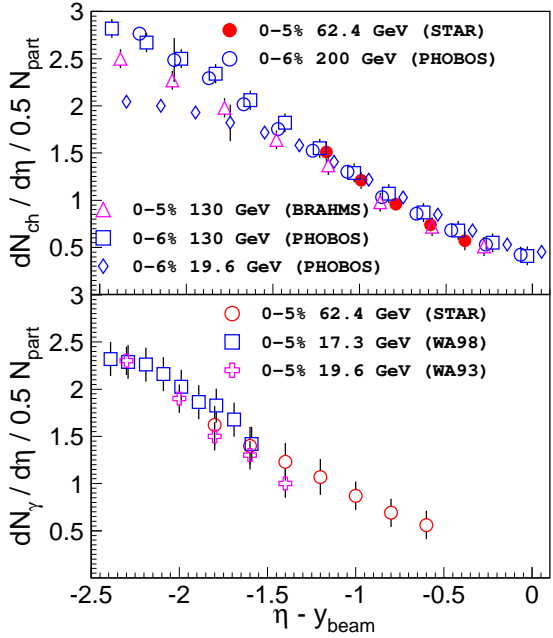}
\caption{(Color Online) The top panel shows the variation of 
$dN_{\mathrm{ch}}/d\eta$ normalized
to  $N_{\mathrm{part}}$  with $\eta$ -- y$_{\mathrm{beam}}$ 
for different collision energies for central collisions. 
The bottom panel shows the variation of 
$dN_{\mathrm{\gamma}}/d\eta$ normalized
to  $N_{\mathrm{part}}$  with $\eta$ -- y$_{\mathrm{beam}}$ 
for different collision energies for central collisions.}
\label{fig9}
\end{center}
\end{figure}
  Continuing our discussion on particle density in $\eta$, we now present
  results on the longitudinal scaling of particle production in 
  heavy ion collisions. It has been observed that the number of 
  charged particles produced per participant pair as a 
  function of $\eta$ -- y$_{\mathrm{beam}}$, 
  where y$_{\mathrm{beam}}$ is the beam rapidity, is independent of 
  beam energy~\cite{phobos,starphoton}. This phenomenon is known as limiting
  fragmentation~\cite{limiting_frag}. Here we present the results on
  energy dependence of limiting fragmentation at 62.4 GeV for charged 
  particles and photons produced in Au + Au collisions. In the subsequent
  sections we discuss the centrality and species dependence of this
  scaling.
 
In Fig.~\ref{fig9} we present the energy dependence of limiting fragmentation
for inclusive charged particles and photons. The charged particle 
pseudorapidity distribution for central (0--5\%) Au + Au collisions  
at $\sqrt{s_{\mathrm{NN}}}$ = 62.4 GeV is compared to the charged particle 
pseudorapidity 
distributions from PHOBOS for central (0--6\%) collisions  
at 19.6 GeV, 130 GeV and 200 GeV~\cite{phobos} and 
charged particle pseudorapidity distribution from BRAHMS for 
central (0--5\%)  collisions  at 130 GeV~\cite{brahms}. 
The photon pseudorapidity distributions 
for central (0--5\%) Au + Au collisions at $\sqrt{s_{\mathrm{NN}}}$ = 62.4 GeV is 
compared with central (0--5\%) photon data for Pb + Pb collisions at 17.3 GeV 
from the WA98 experiment~\cite{wa98_dndy} and 19.6 GeV central (0--5\%) S + Au 
collision data 
from the WA93 experiment~\cite{wa93_dndy}. We observe in Fig.~\ref{fig9} that 
the SPS and RHIC (62.4 GeV) photon results are consistent with each other, 
suggesting that photon production follows an energy independent limiting
fragmentation  behavior. The charged particles at 62.4 GeV also
show an energy independent limiting fragmentation behavior.

\subsection{Centrality dependence of limiting fragmentation}
\begin{figure}
\begin{center}
\includegraphics[scale=0.42]{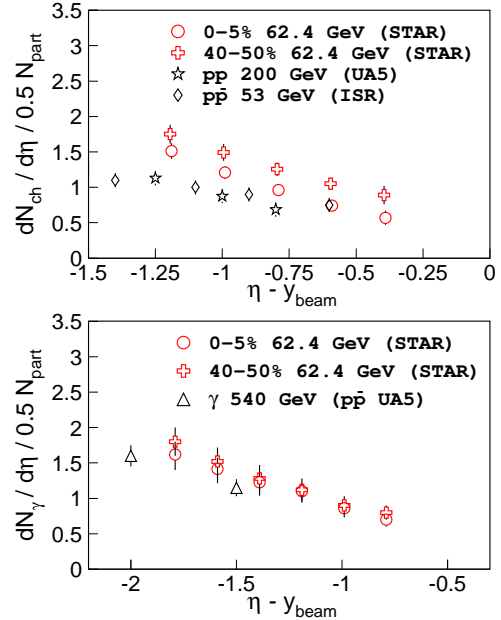}
\caption{ (Color Online)
The top panel shows the variation of 
$dN_{\mathrm{ch}}/d\eta$ normalized
to  $N_{\mathrm{part}}$  with $\eta$ -- y$_{\mathrm{beam}}$ 
for the central and peripheral collisions. Bottom panel shows  the variation of
$dN_{\mathrm{\gamma}}/d\eta$ normalized
to  $N_{\mathrm{part}}$  with $\eta$ -- y$_{\mathrm{beam}}$ 
for the central and peripheral collisions.
Also shown are the charged particle and photon yields in {\it p} + {\it p} 
and {\it p} + {\it $\bar{p}$} collisions.}
\label{fig10}
\end{center}
\end{figure}
 Recently there have been contradictory results reported from inclusive 
 charged particle measurements regarding the centrality dependence of the 
limiting fragmentation 
  behavior. Results from PHOBOS show a centrality dependence~\cite{phobos},
  while those from BRAHMS show a centrality independent 
 behavior~\cite{brahms}. Here we present the
  results on the centrality dependence of limiting fragmentation for
  charged particles and photons at $\sqrt{s_{\mathrm{NN}}}$ = 62.4 GeV.

In Fig.~\ref{fig10} we show the centrality dependence of limiting 
fragmentation for charged particles and photons.
The charged particle pseudorapidity distributions for 0--5\% 
and 40--50\% central Au + Au collisions at 
$\sqrt{s_{\mathrm{NN}}}$ = 62.4 GeV have been compared. We observe, at forward rapidity,
the charged particle yield normalized to the number of participating nucleons 
as a function of $\eta$ -- y$_{\mathrm{beam}}$ is higher for peripheral 
collisions compared to central collisions, whereas within the measured $\eta$
range of 2.3 to 3.7, the photon yield normalized to the number of participating
 nucleons as a function of $\eta$ -- y$_{\mathrm{beam}}$ is found to be 
independent of centrality.
The dependence of limiting fragmentation on the collision system 
is most clearly seen in the comparison between results from heavy 
ion collisions with those from $p$ + $p$ and $p$ + $\bar{p}$ 
collisions~\cite{ua5}. 
We observe in Fig.~\ref{fig10} that the photon results in the forward 
rapidity region from $p \bar{p}$ collisions at $\sqrt{s_{\mathrm{NN}}}$ = 540 GeV 
are in close agreement with the measured photon yield in Au + Au collisions 
at $\sqrt{s_{\mathrm{NN}}}$= 62.4 GeV. However the $p$ + $p$ and $p$ + 
$\bar{p}$ inclusive 
charged particle results are very different from those for 
Au + Au collisions at $\sqrt{s_{\mathrm{NN}}}$= 62.4 GeV. 
It may be mentioned that
the photon yield is dominated by photons from decay of 
$\pi^0$s~\cite{starphoton}. The presented photon results and their 
comparison with
nucleon-nucleon collisions indicate that in 
the $\eta$ region studied, there is apparently a significant charged baryon 
contribution in nucleus-nucleus collisions.
Similar centrality dependent behavior of limiting fragmentation for 
charged particles was also observed by PHOBOS~\cite{phobos}.
The centrality dependence of limiting fragmentation in charged particles 
has been speculated to be due to nuclear remnants and baryon 
stopping~\cite{phobos,baryon_junction}. The centrality independent 
limiting fragmentation for photons has been attributed to mesons being
the dominant source of photon production~\cite{starphoton}.  HIJING 
calculations indicate that about 93--96\% of the 
photons are from $\pi^0$ decays.

\begin{figure}
\begin{center}
\includegraphics[scale=0.42]{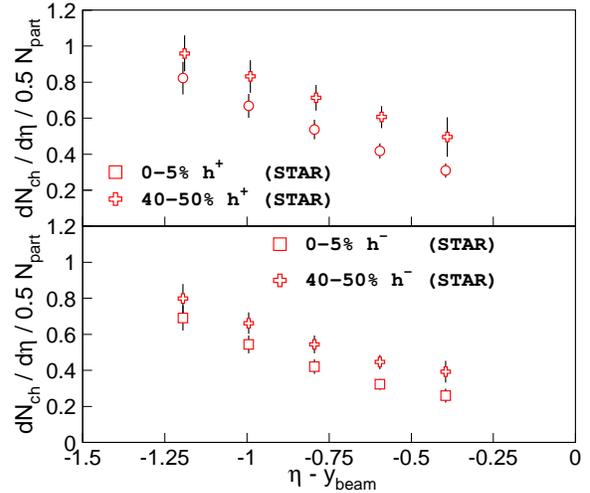}
\caption{ (Color Online)
Variation of $dN_{\mathrm{ch}}/d\eta$ normalized
to  $N_{\mathrm{part}}$  with $\eta$ -- y$_{\mathrm{beam}}$ 
for central and peripheral collisions for positively
charged hadrons ($h^{+}$) and negatively charged hadrons ($h^{-}$).
}
\label{fig11}
\end{center}
\end{figure}
In order to understand the role of nuclear remnants and baryon stopping
in the observed centrality dependent behavior of limiting fragmentation
of charged particles, we have studied the limiting fragmentation for
positively and negatively charged hadrons separately. The contribution from 
protons coming from beam remnants can be understood by studying the 
limiting fragmentation of positively charged hadrons. In Fig.~\ref{fig11}
we have plotted $\frac{d N_{\mathrm{ch}}}{d \eta}$ normalized
to the number of participating  nucleons for 40--50\% 
and for 0--5\% central collisions for positively ($h^{+}$) and
negatively charged ($h^{-}$) hadrons. 
In addition to the systematic errors discussed earlier, and shown in the
figure, there is an error due to the uncertainty in the charge 
determination. The uncertainty has been studied
by embedding charged 
Monte Carlo tracks into real data and then following the
full reconstruction chain. This error was obtained
as a function of $\eta$. It is defined as the ratio of the total number of 
embedded charged tracks whose charge has been reconstructed incorrectly,
to the total number of charged tracks embedded. The error in charge 
determination was found to
increase from 2\% at $\eta$ = 2.9 to 15\% at $\eta$ = 3.9.
We find that both $h^{+}$ and $h^{-}$ show a centrality dependent 
limiting fragmentation behavior. When compared to the centrality independent
limiting fragmentation behavior for photons (Fig.~\ref{fig10}) and to 
results from nucleon-nucleon collisions 
(Fig.~\ref{fig10}), our measurements indicate that baryon transport 
at forward rapidity also plays an important role in the observed 
centrality dependent behavior of limiting fragmentation for charged particles.
We find that the ratio for yields of $h^{+}$ from peripheral
to central 
collisions  increases from 1.17$\pm$0.06 at $\eta$ = 3.0
to 1.61$\pm$0.07 at $\eta$ = 3.8 (closer to beam rapidity). 
The values for $h^{-}$  are 1.16$\pm$0.06 at 
$\eta$ = 3.0 and  1.51$\pm$0.07 at $\eta$ = 3.8.
From these values we find that the increase in the ratio 
with $\eta$ seems to be somewhat weaker for $h^{-}$ compared to $h^{+}$. 
However, within the systematic errors, it is difficult 
to conclude on the role of the beam remnants (beam protons in $h^{+}$) in the 
centrality dependent behavior of limiting fragmentation for charged 
particles at forward rapidity.

\begin{figure}
\begin{center}
\includegraphics[scale=0.42]{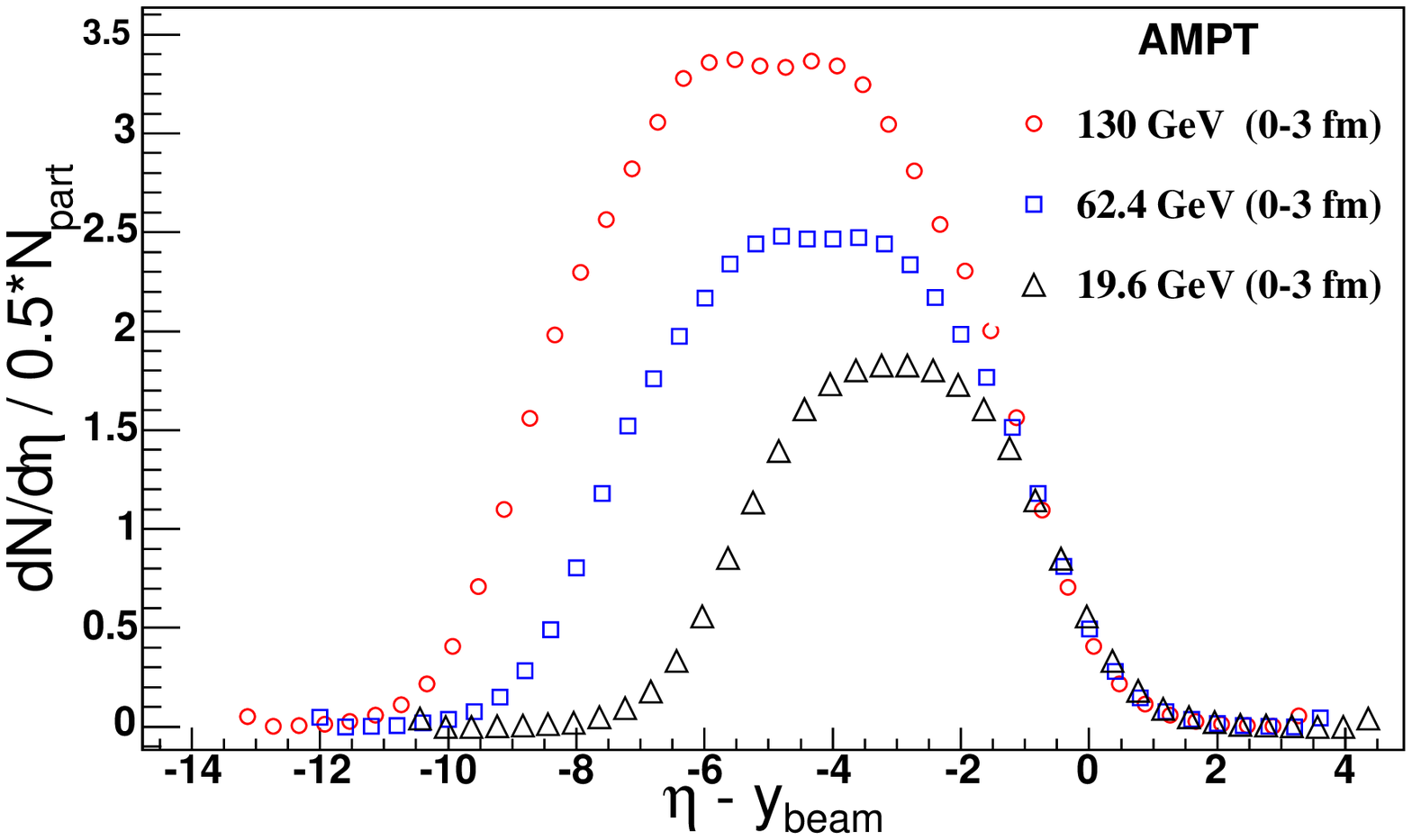}
\includegraphics[scale=0.42]{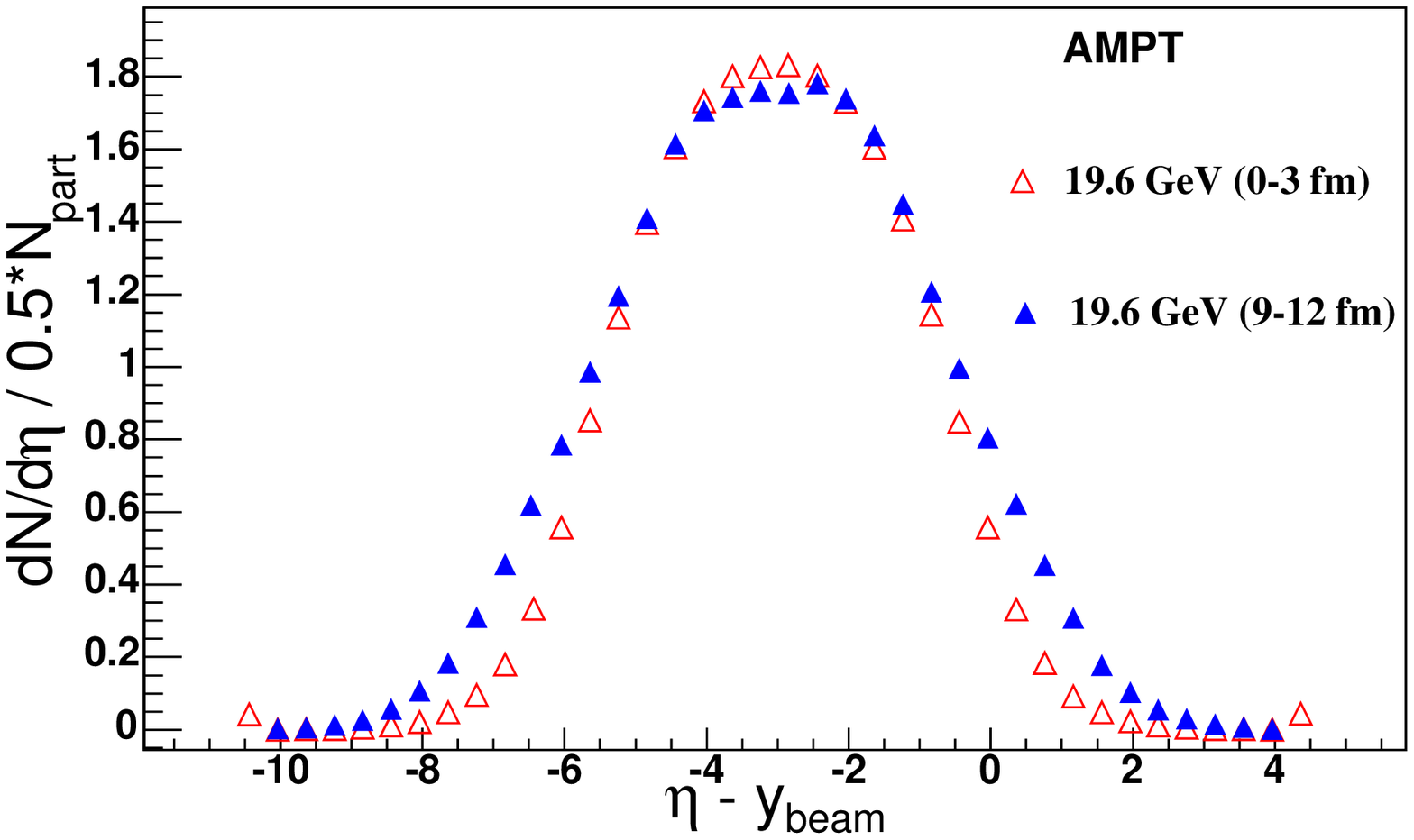}
\caption{ (Color Online)
Variation of $dN_{\mathrm{ch}}/d\eta$ normalized
to  $N_{\mathrm{part}}$  with $\eta$ -- y$_{\mathrm{beam}}$ 
from AMPT model~\cite{ampt} calculations 
for various center of mass energies in central collisions (top panel)
and central and peripheral collisions at $\sqrt{s_{\mathrm{NN}}}$= 19.6 GeV
(bottom panel).
}
\label{fig13}
\end{center}
\end{figure}
Energy and centrality dependence of limiting fragmentation for 
charged particles can be a test for particle production models. 
We have observed that particle production models such as HIJING and AMPT 
are not able to describe fully the $\eta$ distribution of charged 
particles at forward rapidity. However, it is interesting to 
investigate whether they can qualitatively reproduce the limiting fragmentation 
features of experimental data. Our calculations show that in the HIJING 
and AMPT models the charged particles show energy independent 
limiting fragmentation. The centrality dependent behavior of limiting
fragmentation for charged particles is more clearly observed
in the AMPT model than in HIJING. In  Fig.~\ref{fig13} we only show 
the results from the AMPT model. These results 
are for the $\sqrt{s_{\mathrm{NN}}}$
values of 19.6 GeV, 62.4 GeV and 130 GeV Au + Au collisions 
at 0--3 fm and 9--12 fm impact parameter. For the centrality dependence 
we only show the results for $\sqrt{s_{\mathrm{NN}}}$= 19.6 GeV, the energy 
at which the centrality dependent effect is most prominent in 
the data~\cite{phobos}. 
The AMPT model has qualitative limiting fragmentation 
features similar to those of experimental data (shown in Fig.~\ref{fig10}). 
We find in the model that the central yields, when normalized to
number of particpating nucleons, are also lower than the corresponding 
peripheral yields at forward rapidity when $\eta$ is shifted by the 
beam rapidity.

\subsection{Identified particle limiting fragmentation}
\begin{figure}
\begin{center}
\includegraphics[scale=0.5]{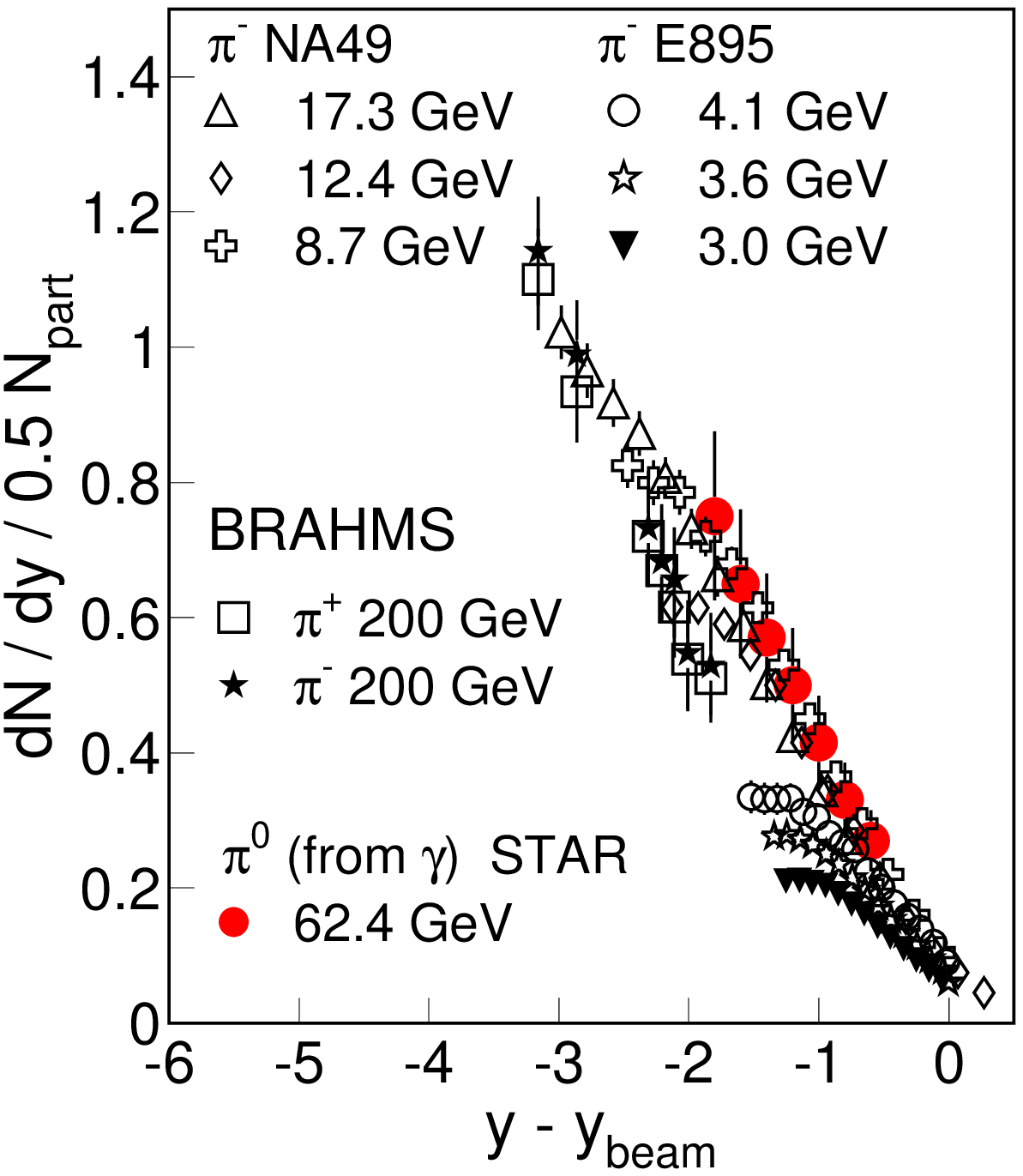}
\includegraphics[scale=0.5]{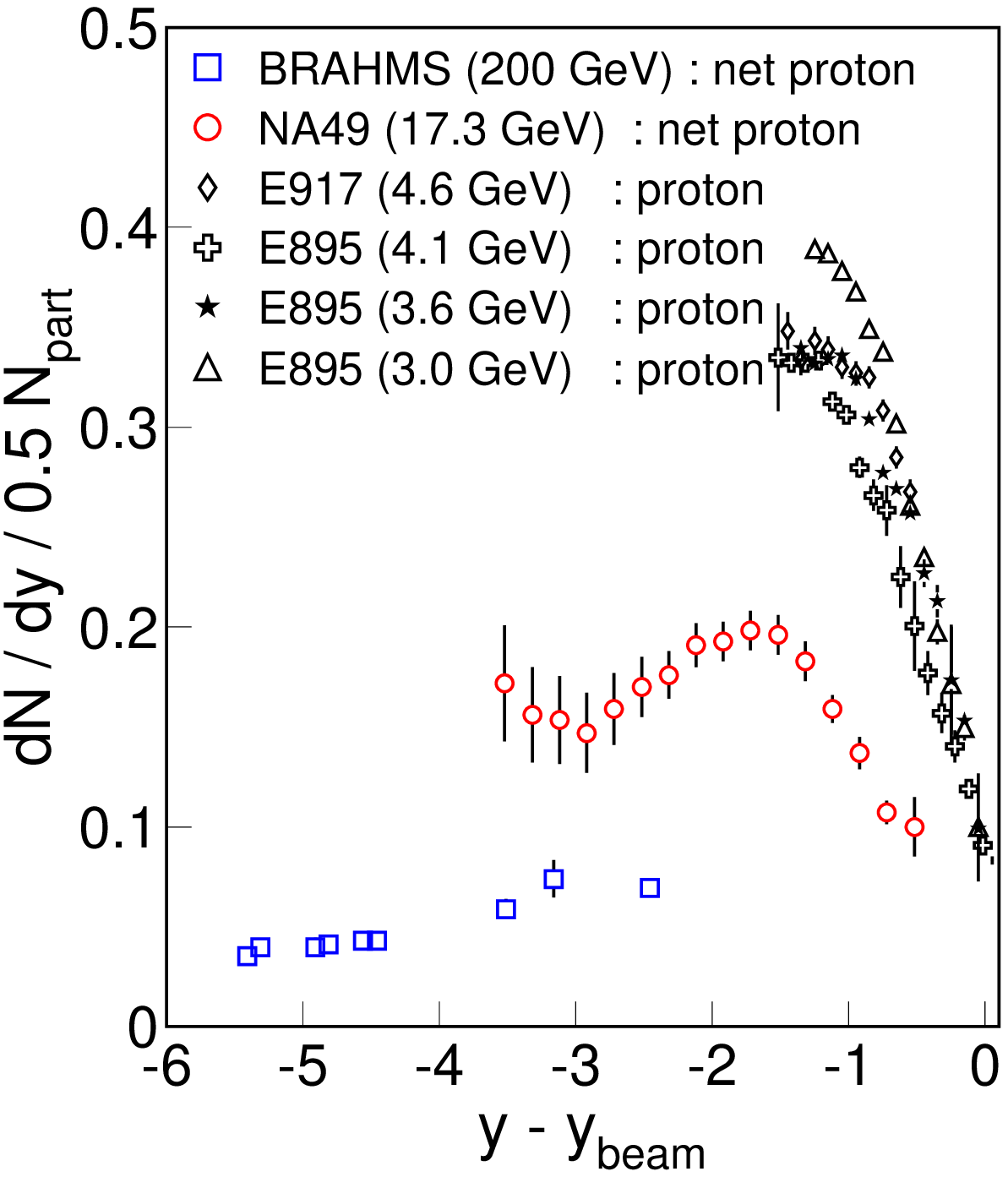}
\caption{ (Color Online)
The top panel shows the variation of pion rapidity density
normalized to $N_{\mathrm{part}}$ with
y -- y$_{\mathrm{beam}}$ for central collisions at various collision energies.
Also shown is the estimated $dN_{\mathrm \pi^{0}}/dy$
obtained from $dN_{\mathrm{\gamma}}/dy$ normalized to $N_{\mathrm{part}}$.
The bottom panel shows the variation of net proton rapidity density
normalized to  $N_{\mathrm{part}}$ with y -- y$_{\mathrm{beam}}$ 
for central collisions at various collision energies. 
}
\label{fig12}
\end{center}
\end{figure}
The observation of centrality dependent and energy independent limiting 
fragmentation for inclusive charged particles, along with the centrality
and energy independent limiting fragmentation for photons (presented 
in previous sections), motivates us 
to study the limiting fragmentation of identified particles.

The top panel in Fig.~\ref{fig12} shows the charged pion rapidity density 
in central Au + Au collisions at RHIC~\cite{rhic_pion}, Pb + Pb collisions at the 
SPS~\cite{na49_pion} and Au + Au collisions at AGS~\cite{ags}. Also 
shown is the estimated $\pi^{0}$  rapidity density from the present 
measurement of the photon rapidity density at $\sqrt{s_{\mathrm{NN}}}$= 62.4 GeV, 
all as a function of y--y$_{\mathrm{beam}}$~\cite{starphoton}.  
We obtained the ratio of the photon to $\pi^0$ yields from HIJING.
This ratio is used to estimate the $\pi^0$ yield from the 
measured photon yield. The results indicate
that the pion production in heavy ion collisions in the fragmentation
region agrees with the energy independent limiting fragmentation picture. 

The bottom panel of  Fig.~\ref{fig12} shows the net proton ($p$ - $\bar{p}$) rapidity density
in central Au + Au collisions at RHIC~\cite{brahms_proton} energies and 
Pb + Pb collisions at SPS~\cite{na49_proton} energies. 
For AGS energies~\cite{ags,ags_proton} we plot only the
proton rapidity density in Au + Au collisions. Since the anti-proton yields 
are very low ($\bar{p}/p$ $\sim$ 2 $\times$ $10^{-4}$ 
at top AGS energy), the proton rapidity density
reflects the net proton rapidity distribution.
The net protons violate the energy dependence of 
limiting fragmentation. These results show that baryons and mesons 
differ in the energy dependence of limiting fragmentation.
The results for identified particles, along with the centrality 
dependence of limiting fragmentation for inclusive 
charged hadrons, and the centrality 
independence of limiting fragmentation for identified mesons,
shows that the baryon transport in heavy ion 
collisions plays an important role in particle production at 
forward rapidity. The results also show that although
baryon stopping is different in different collision systems,
the pions produced at forward rapidity are not affected by
baryon transport. The limiting fragmentation study for net protons 
may also indicate the validity of a baryon junction 
picture~\cite{baryon_junction}. 
If the baryon numbers are carried by the valence quarks,
then at forward rapidity the baryons should also follow
an energy independent limiting fragmentation behavior,
like pions (originating from valence quarks).
This may indicate that the baryon number is not carried by
the valence quark, which is suggested in the 
baryon junction picture, where the baryon number 
resides in a non-perturbative configuration of gluon fields,  
rather than in the valence  quarks.

\section{Summary}
In summary, we have presented charged particle and photon multiplicity
measurements at RHIC in the pseudorapidity regions 
2.9 $\le$ $\eta$  $\le$ 3.9 and 2.3 $\le$ $\eta$  $\le$ 3.7, respectively.
The pseudorapidity distributions of charged particles and photons
for Au + Au collisions at $\sqrt{s_{\mathrm{NN}}}$= 62.4 GeV have been 
obtained for 
various centrality classes and compared to results from different models. 
Charged particle and photon production normalized to the number of 
participating nucleon pairs and to the number of binary collisions 
has been studied. The photon multiplicity, within the 
systematic errors, seems to scale with the number of participating nucleons,
while the charged particle multiplicity does not.
Both the photon and charged particle 
production at forward rapidity do not scale with number of binary 
collisions. This indicates that the particle production at forward 
rapidity is not dominated by a contribution from hard processes. Charged 
particle and photon distributions at $\sqrt{s_{\mathrm{NN}}}$ = 62.4 GeV are 
both observed to be consistent with the energy independent 
limiting fragmentation scenario. 
The photon production is observed to follow a 
centrality independent limiting fragmentation scenario,
while the charged particles follow a centrality dependent 
behavior. Comparison of the pseudorapidity distributions of positively
charged particles, negatively charged particles, pions, and 
distributions from {\it p} + {\it p} collisions, indicate that 
the baryons are responsible
for the centrality dependent limiting fragmentation behavior of 
charged particles. 
The study of limiting fragmentation for pions and net 
protons show that mesons follow energy independent limiting fragmentation, 
whereas baryons do not. \\

\begin{acknowledgments}
We thank the RHIC Operations Group and RCF at BNL, and the
NERSC Center at LBNL for their support. This work was supported
in part by the HENP Divisions of the Office of Science of the U.S.
DOE; the U.S. NSF; the BMBF of Germany; IN2P3, RA, RPL, and
EMN of France; EPSRC of the United Kingdom; FAPESP of Brazil;
the Russian Ministry of Science and Technology; the Ministry of
Education and the NNSFC of China; SFOM of the Czech Republic,
FOM and UU of the Netherlands, DAE, DST, and CSIR of the Government
of India; the Swiss NSF; the Polish State Committee for Scientific
Research; STAA of Slovakia, and the Korea Sci. \& Eng. Foundation.
We acknowledge the help of CERN for use of GASSIPLEX chips in the PMD readout.
We thank  S.~Jeon for providing us the LEXUS results for comparison and
G.~Veres for providing us the NA49 data on net protons.
\end{acknowledgments}


\end{document}